\documentclass[10pt,a4paper,english,journal]{IEEEtran}
\usepackage[T1]{fontenc}
\usepackage[latin9]{inputenc}
\usepackage{graphicx}

\makeatletter

\pdfpageheight\paperheight
\pdfpagewidth\paperwidth

\providecommand{\tabularnewline}{\\}

\usepackage{xcolor,soul}
\sethlcolor{lightgray}

\makeatother

\usepackage{babel}
\begin{document}
\title{Power Saving Techniques in 3GPP 5G New Radio: A Comprehensive Latency
and Reliability Analysis }
\author{\IEEEauthorblockN{Ali A. Esswie\textit{ }\\
Cellullar Standards Technical Lead, Advanced Air Interface, Future Wireless, InterDigital Communications.
ali.esswie@interdigital.com}}

\maketitle
$\pagenumbering{gobble}$
\begin{abstract}
Energy efficiency is critical for future sustainable cellular systems.
Power saving optimization has been a key part of the fifth generation
(5G) new radio specifications. For 5G-advanced and future 6G, with
the anticipation of a trillion internet of things (IoTs) devices with
non-rechargeable or low-density batteries, device power efficiency
is rather essential. There are numerous contributions from industry
and academia which present the potential power saving gains of the
various 5G power saving techniques; however, there is a lack of art
on the performance cost paid to achieve such power saving gains. Therefore,
this paper presents a comprehensive evaluation of the radio latency
and reliability cost, which is lost due to a certain 5G new radio
power saving feature. A thorough review of the state-of-the-art 5G
power saving techniques is introduced. Extensive system level simulations
are performed to evaluate the latency and reliability cost of the
considered power saving features. The paper offers valuable recommendations
for supporting power-efficient latency-critical traffic for beyond
5G-advanced systems.

\end{abstract}

\section{Introduction}

The fifth generation (5G) new radio (NR) is driven by the deterministic
ultra-low latency use-cases, i.e., the ultra-reliable and low-latency
communications (URLLC) {[}1{]}. Those support a wide variety of industrial
internet of things (IIoTs) deployments of intelligent sensors, actuators,
robots, and machines {[}2{]}. To support the critical industrial applications,
the majority of the IIoT URLLC communications demand a stringent set
of the radio latency and reliability, which is comparable to that
is of the Ethernet-based communications, i.e., a one-way radio latency
of 1-5 ms with a success probability of 99.999\%. 

For beyond 5G-advanced systems, i.e., beyond 3GPP release-20 (6G),
it is envisioned to deploy a trillion IIoT devices with the requirement
of a prolonged battery lifetime, i.e., beyond 10-year of operation
{[}3{]}. Furthermore, it has been seen, from field deployments and
lab tests, that such battery requirement is infeasible in practice
and accordingly, the typical IIoT device lifetime becomes much more
longer than that is of the battery. In particular, those battery targets
assume the IIoT devices are deep sleeping for extended periods of
time and only waking up when there is a payload for transmission.
This is not appropriate for mobile terminated services, where devices
need to at least periodically search for potential incoming traffic.
Thus, despite the battery lifetime assumptions, it leads to millions
of IIoTs devices removed and manually recharged or completely thrown-away
per day, which imposes a significant expenditure overhead and hazardous
environmental waste, respectively {[}4{]}. 

Those power efficiency challenges have been well reflected by a variety
of state-of-the-art contributions from 3GPP partners and academia.
Specifically, energy efficient radio communications are recognized
as a vital design target of the upcoming 6G cellular networks. For
5G new radio, there has been an extensive work on device power saving
procedures {[}5{]}, including power optimizations for idle, inactive
and connected modes. On another side, and away from the power efficiency
targets, the feasibility of the URLLC latency and reliability requirements
is investigated for macro deployments with frequency division duplex
(FDD) {[}6{]}, time division duplex (TDD) {[}7{]}, and for industrial
factory TDD deployment {[}2{]}, respectively. Furthermore, smarter
scheduling algorithms {[}8, 9{]} are vital to achieve the challenging
URLLC targets, particularly in coexistence deployments with various
quality of service (QoS) classes on the same spectrum. However, there
is lack in the state-of-the-art literature on the performance cost,
i.e., radio latency and reliability, paid in order to fulfill a power
efficient operation at the device side. 

In this work, the state-of-the-art 3GPP power saving techniques have
been extensively investigated and reviewed. Those include the power
efficient RRC inactive state, the flexible discontinuous reception
(DRX), the cross-slot scheduling, the paging early procedures, small
data transmission without RRC state transition, and dynamic reference
signal sharing, respectively. Accordingly, an extensive set of realistic
system-level simulations are carried out to evaluate the achievable
latency and reliability performance for each of the considered power
saving feature. The paper offers valuable recommendations on the current
power efficiency challenges towards 6G systems. 

This paper is organized as follows. Section II briefly presents the
system model of this work while Section III surveys the state-of-the-art
power saving techniques until the current 3GPP release-17. Section
IV introduces the simulation results. Finally, the paper is concluded
in Section V. 

\section{System Model}

An indoor industrial factory deployment is considered in this work,
where $C$ cells are deployed to offer coverage to the factory area
and each serves $K$ uniformly-distributed UEs. Both downlink and
uplink directions are separately considered. The URLLC-alike FT3 traffic
is considered with small and sporadic payload transmissions, where
the offered traffic load per cell is calculated in line with {[}9{]}. 

The 3GPP methodology for URLLC system-level simulations is followed
in this work. UEs are dynamically multiplexed using the orthogonal
frequency division multiple access (OFDMA), with each physical resource
block (PRB) of 12 successive sub-carriers. A short transmission time
interval (TTI) duration of 4-OFDM symbols is considered alongside
with 30 KHz sub-carrier spacing for faster URLLC transmissions. The
performed system-level simulations include the majority of Layer 1
and 2 functionalities, in a realistic setup, such as dynamic UE scheduling
(using proportional fair), dynamic hybrid automatic repeat request
(HARQ) re-transmission and combining, indoor industrial factory channel
modeling, and dynamic link adaptation, respectively. The major performance
key indicator of this work is the outage latency {[}7{]}, where it
denotes the achievable radio latency of each transmitted payload until
it is successfully decoded by its intended receiver, for a certain
perceived outage probability. That is, including the inter-UE scheduling
queuing delay, cell and UE processing delays, HARQ re-transmission
delay, RRC state transition from IDLE/Inactive modes to connection
mode, and paging detection delay. 

\section{5G New Radio: Power Saving Techniques}

\subsection{RRC Inactive State}

\begin{figure}
\begin{centering}
\includegraphics[scale=0.5]{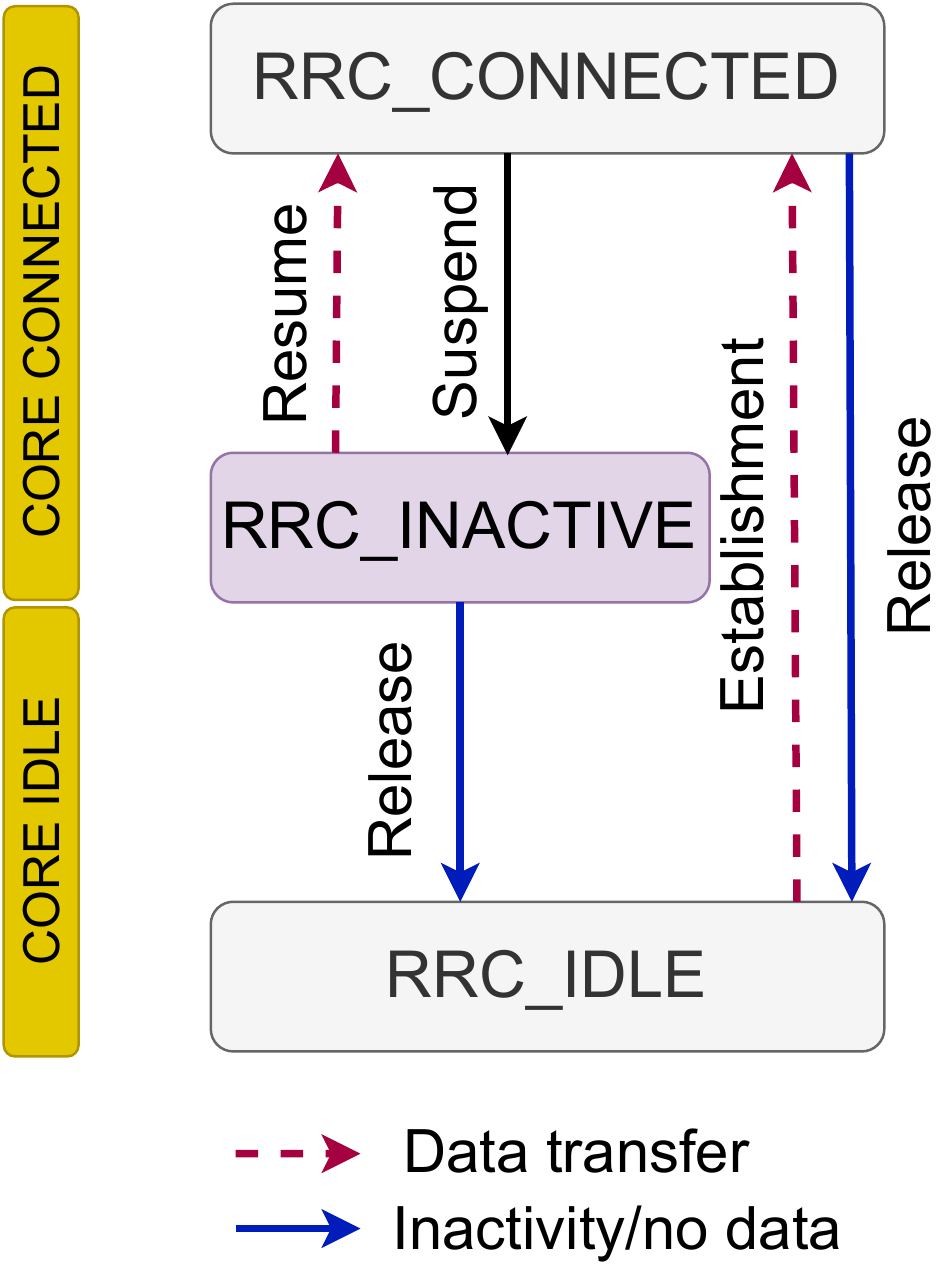}
\par\end{centering}
\centering{}Fig. 1. RRC state transition in 5G new radio.
\end{figure}

The RRC Inactive state {[}5{]} has been introduced since 3GPP releasse-15,
with the objective of reducing the UE power consumption and the connection
establishment control overhead, respectively. With the RRC Inactive
state, the UE core-network context information is kept alive at the
last known gNB, i.e., anchor gNB. Therefore, as depicted by Fig. 1,
when the UE transitions back to RRC connected mode, for payload transmission
or reception, the current selected gNB acquires the UE context from
the anchor gNB, and the RRC connection is established accordingly.
The RRC Inactive state therefore enables a faster RRC connection establishment
without the need to establish the core-network connection and respective
security keys. This is useful for URLLC services for which sporadic
and frequent traffic arrivals are foreseen such that UEs avoid the
RRC connection establishment overhead.

Upon data inactivity, The gNB triggers an RRC suspend command to the
UE and the UE enters the RRC Inactive state, with the core-network
context information kept active. For both the RRC IDLE/Inactive states,
the UE monitors the configured paging occasions. When a true paging
indication is detected and decoded, the RRC IDLE mode UE must transition
to the RRC connected mode before the payload transmission or reception;
however, the RRC Inactive mode UE may transmit or receive a small
payload without transitioning to RRC connected mode.

\subsection{Flexible Discontinuous Reception (DRX) }

DRX is a legacy measure for connected-mode UEs to sleep and shut down
their transceiver chains for extended periods of time, accordingly,
avoiding the excessive battery consumption. A DRX cycle is defined
by a periodic set of wake-up times over which the UE shall wake-up/turn
on its receiver, and hence, attempt decoding the configured PDCCH
control channels, for detecting a possible scheduling grant. There
are long and short cycled DRX. The short DRX cycle spans from 2 ms
to 640 ms while the long DRX cycle is from 10 ms to 10.24 s. Clearly,
the long DRX cycle offers better UE power saving gain, at the expense
of longer radio latency, i.e., the device is not reachable during
the DRX OFF periods.

Therefore, the UE, according to its own implementation, decides the
sleep state which it shall trigger between the DRX ON opportunities.
As part of 3GPP release-17, there are three sleep states defined for
5G devices as: deep sleep, light sleep, and micro sleep. Each state
denotes that UE shuts down certain components of its RF chain. For
instance, a UE within the deep sleep state implies that UE RF chain
is completely shutdown and the UE can not monitor or receive control
or data channels. Thus, to transition from one state to another, the
UE consumes an amount of power and processing delay until it fully
transitions to the required state. For example, to transition from
deep sleep to active state, the UE consumes an average of 20 ms until
it is fully available for reception {[}5{]}. 

Thus, as shown by Fig. 2, 3GPP has introduced an extended inactivity
timer {[}5{]} as part of the connected-mode DRX operation. Upon detecting
a control channel grant during the DRX ON period, the UE stays active
for the duration of the configured inactivity timer in order to continue
monitoring the control channel for possible following grants of further
incoming and/or buffered traffic payload at the gNB. This way, the
UE does not transition quickly between states, and hence, avoiding
the power and delay overhead of the state transition. 

\begin{figure}
\begin{centering}
\includegraphics[scale=0.48]{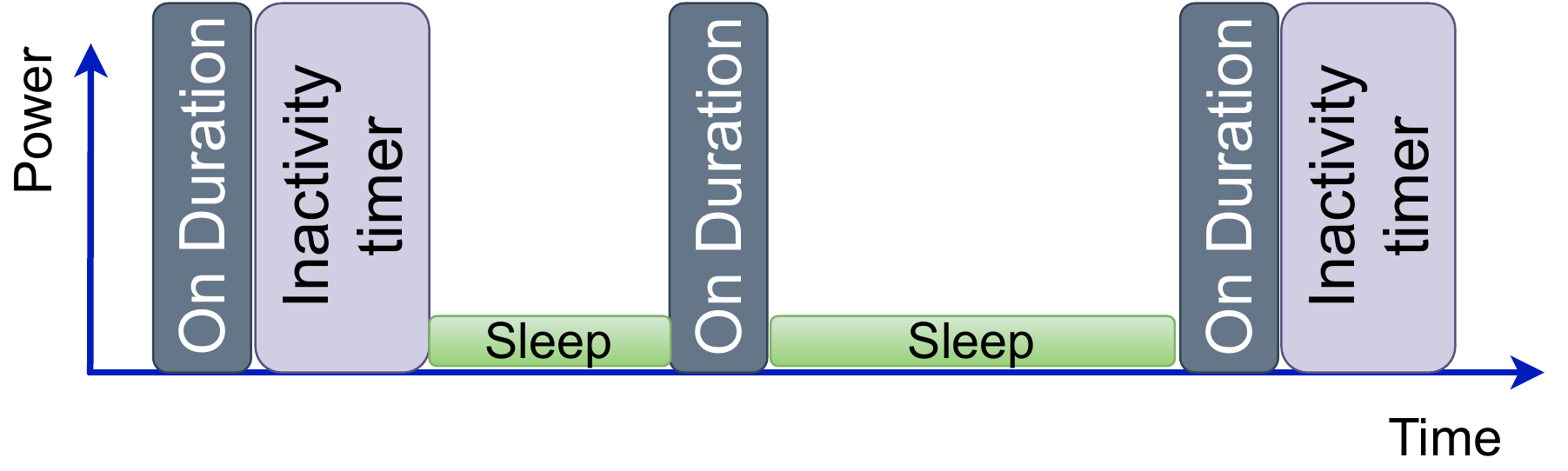}
\par\end{centering}
\centering{}Fig. 2. Connected-DRX (C-DRX) for power critical 5G devices.
\end{figure}

\subsection{Cross-Slot Scheduling}

The 5G new radio, since the early release-15 development, has introduced
a flexible resource allocation for latency-critical QoS UEs. That
is, by dynamically allocating the control and data channels within
a sub-slot duration. For instance, a UE may receive the resource grant
in a slot while the actual allocated data resources are within the
same slot, and possibly, concurrent with the control channel resources.
Thus, UEs, which are configured to monitor a certain PDCCH control
channel search space, will always receive and buffer the concurrent
PDSCH data resource within the same slot, in case they have an immediate
grant, and which they become aware of after fully decoding and processing
the PDCCH channel resources. In case UEs do not detect active resource
allocations, they flush/erase the buffered PDSCH payload. However,
the buffering requirement of the data channels requires a larger buffer
size at the UEs, which accordingly, is not suitable for Reduced Capability
UEs, and hence, imposes an unnecessary power consumption for latency-non-critical
QoS payload. 

Therefore, as depicted by Fig. 3, a cross-slot scheduling solution
is introduced {[}5{]}. UEs are configured, by high level signaling
(RRC, SIB), with the minimum scheduling offset. This offset implies
the minimum delay, in number of symbols and/or slots, between receiving
a resource grant and the actual data resources the UEs should expect.
This removes the need for UEs to buffer concurrent data payload during
the decoding of the PDCCH control channel, and hence, offering a clear
UE power saving gain and relaxing the requirement on the buffer size
of the UEs. The gNB may enable dynamic cross-slot scheduling where
UEs dynamically identify, from decoding the control channel, on which
slots they should expect receiving the actual data payload. The latter
dynamic procedure still fulfills the configured minimum scheduling
offset such as UEs completely avoid the unnecessary data buffering. 

\begin{figure}
\begin{centering}
\includegraphics[scale=0.48]{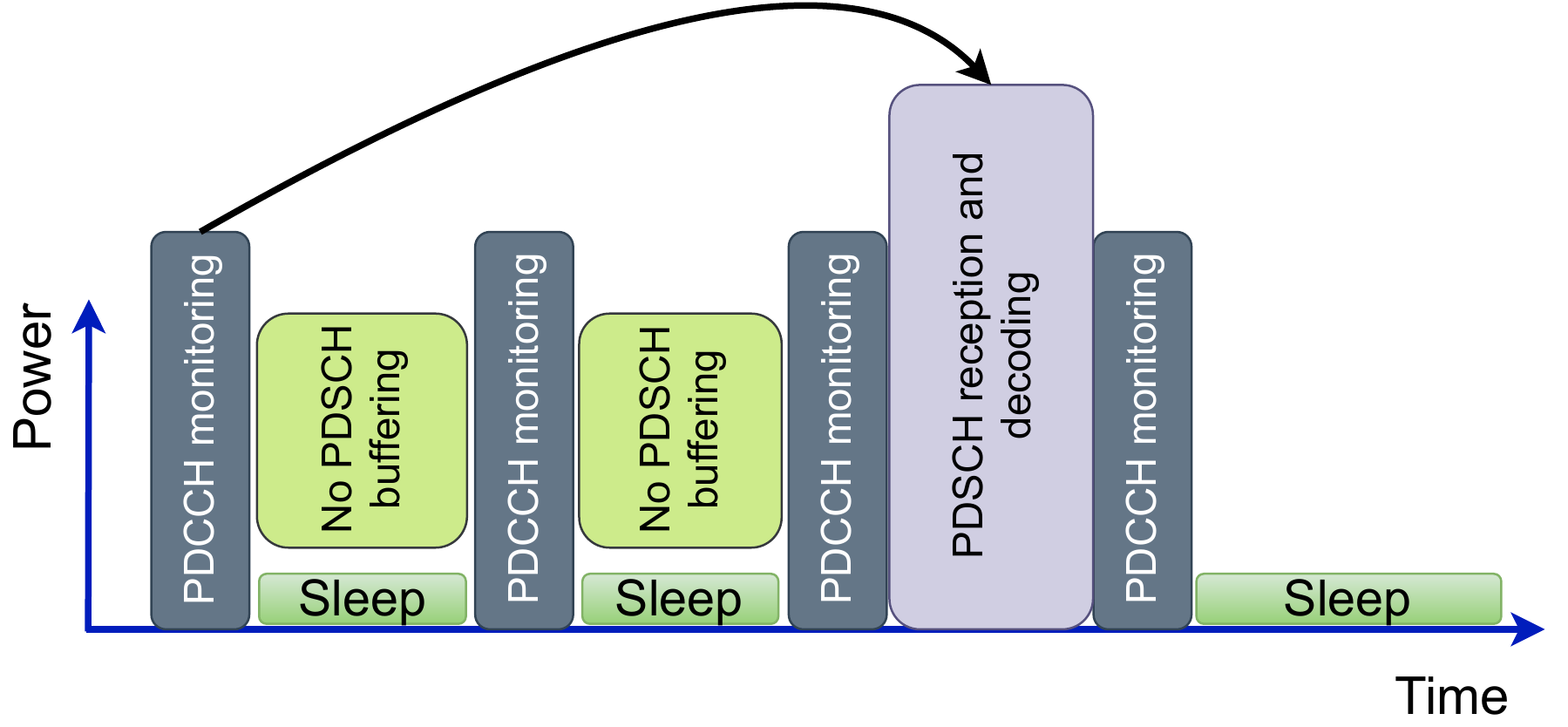}
\par\end{centering}
\centering{}Fig. 3. Cross-slot scheduling and resource allocation.
\end{figure}

\begin{figure}
\begin{centering}
\includegraphics[scale=0.47]{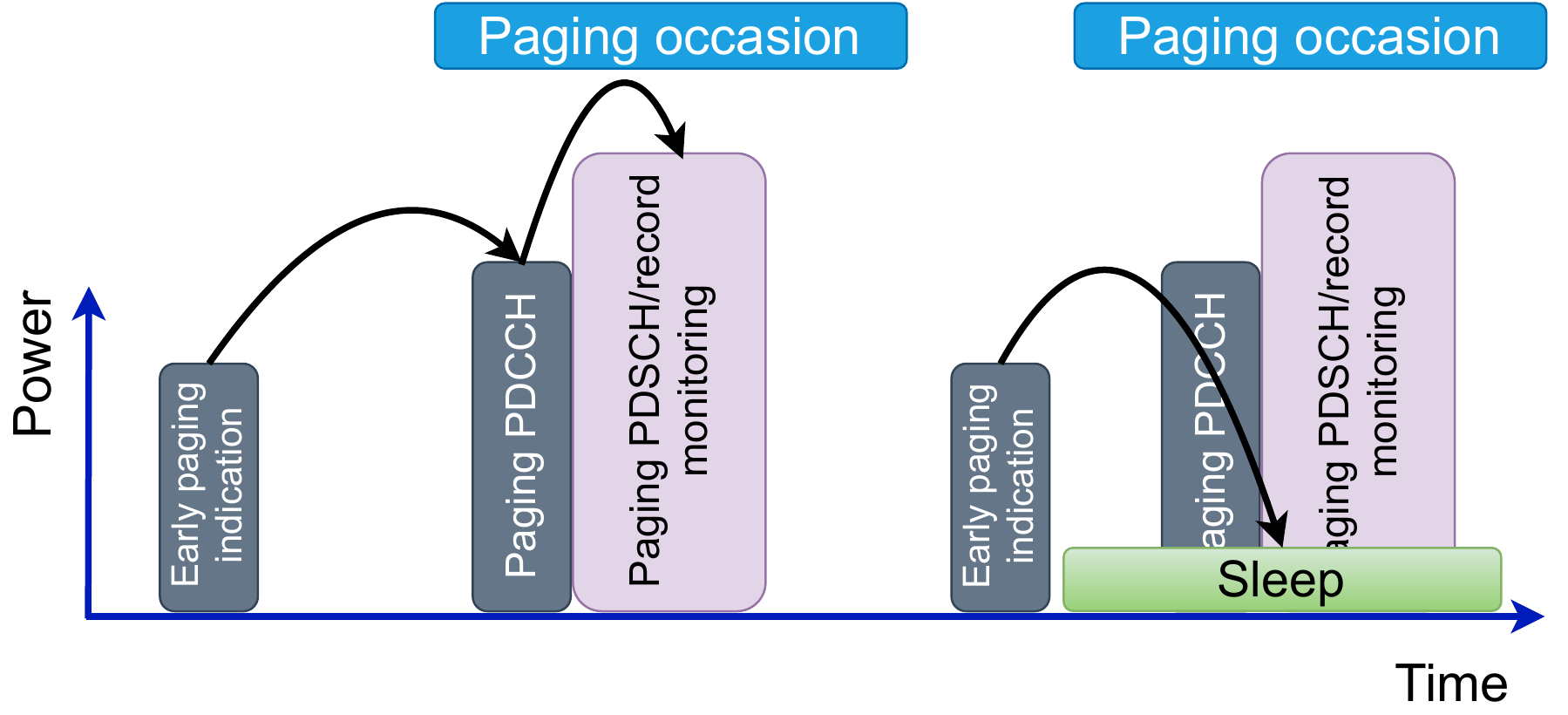}
\par\end{centering}
\centering{}Fig.4. Power-optimized early paging indication.
\end{figure}

\subsection{Early Paging Indication}

For mobile terminated communications, the UEs need to periodically
monitor the configured paging channels to detect upcoming traffic.
Thus, a set of periodic paging occasions (POs) is configured, where
the IDLE/Inactive UEs monitor various POs. A PO is defined by a PDCCH
search space and an associated paging PDSCH record. That is, UEs monitor
the PDCCH search space of the configured PO, and in case there is
a paging indication present, the UEs receive and decode the paging
record (PDSCH) and become aware of the listed identifiers, e.g., I-RNTIs,
of the UEs which are actually paged. Accordingly, paged UEs trigger
the connection establishment procedure while other UEs transition
back to the deep sleep state. The major challenge is that all IDLE/Inactive
UEs must monitor and decode the paging PDCCH and PDSCH even though
if they are not actually paged, which consumes unnecessary power of
the non-paged UEs.

Thus, there have been several paging enhancements during release-17.
As presented by Fig. 4, the control channel of an early paging indication
(EPI) is introduced {[}5{]}. The EPI implies a limited-size downlink
control information (DCI) search space or a sequence, transmitted
from the gNB prior to each PO. IDLE/Inactive UEs monitor the search
space of the EPI, and upon detection of a present EPI indication,
the UEs monitor the next PO. Otherwise, the UEs /deep sleep and skip
detecting the PO. The achievable power saving gain is due to the more
limited EPI search space, compared to the actual paging PDCCH. Hence,
the EPI reduces the number of unneeded PO decoding, i.e., reducing
paging false alarms, for UEs which are not paged. Furthermore, the
EPI DCI or sequence could be defined for a certain group of IDLE/Inactive
UEs. Particularly, IDLE/Inactive UEs are sub-grouped in several paging
groups, by several introduced grouping means, and the EPI DCI is scrambled
in a group-specific manner. Thus, when an IDLE/inactive mode UE calculates
a wrong cyclic redundancy check (CRC) after decoding the EPI DCI with
its own paging-group scrambling code, it assumes that the transmitted
EPI is meant for one or more of the other paging groups, and accordingly
skips the PO, leading to a further reduction of the paging false alarms.

\subsection{Paging-Specific Assistance Reference Signals}

For IDLE/Inactive UEs to decode the PO, UEs need to be first synchronized
with the RAN interface. Due to the long sleep periods among each two
successive POs, the IDLE/Inactive are likely to be out of the RAN
synchronization. Thus, as shown by the top schematic in Fig. 5, IDLE/Inactive
UEs must wake up to detect several synchronization signal blocks (SSBs)
prior to each PO, to retain RAN synchronization. The exact number
of SSBs required for paging pre-synchronization depends on the signal-to-interference-noise
ratio (SINR) of the UE. For instance, UEs with good SINRs require
a single SSB detection while UEs of low SINRs demand at least 3 SSBs
prior the PO. For the latter case, the IDLE/Inactive UEs must wake
up 80 ms before the PO, assuming a standard 20-ms SSB periodicity,
leading to a significant power consumption. 

Hence, paging assisting reference signals (RS) have been proposed.
This denotes that gNB may transmit one or more of the IDLE/Inactive-specific
RS sequences slightly before the POs such that IDLE/Inactive UEs wake
up for shorter periods of time. That is, instead of waking up for
detecting several SSBs with the longer periodicity, as shown by Fig.
5. The presence of the paging-specific RSs is signaled to the IDLE/Inactive
mode UEs such as they can reliably skip detecting the needed SSBs.
However, such procedure increases the network overhead over the paging
bandwidth part, i.e., default bandwidth part, due to the almost-always-ON
RS transmission (LTE-alike). To avoid such overhead, connected-mode
RSs (TRS, CSI-RS), which are naturally available for ongoing connected-mode
traffic prior to the PO, can be shared to IDLE/Inactive mode UEs as
well, hence, reducing the number of paging dedicated RS transmissions.
However, as the connected-mode RSs are only available when there is
an ongoing traffic, it is a challenging task to rapidly signal IDLE/Inactive
UEs with their presence.

\begin{figure}
\begin{centering}
\includegraphics[scale=0.47]{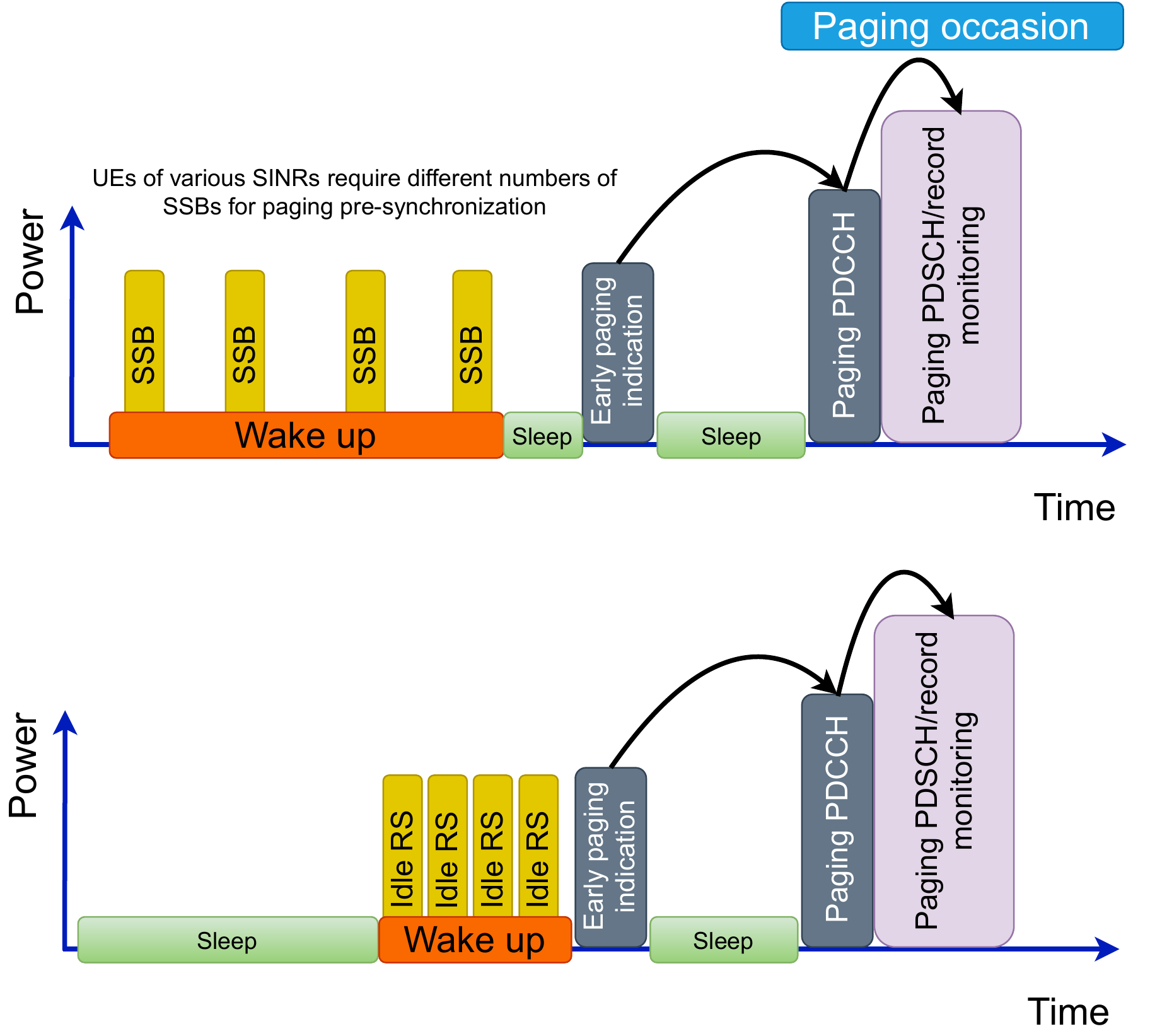}
\par\end{centering}
\centering{}Fig. 5. Idle/inactive mode assisting reference signal.
\end{figure}

\subsection{Inactive Uplink Small Data Transmissions}

For IDLE/Inactive UEs to transmit an uplink payload, they first need
to transition to the RRC connected state by performing the random
access procedure (RACH), and the RRC connection establishment procedure,
respectively. For URLLC services, the useful payload is typically
small-sized, e.g., 50 bytes, while the needed RRC signaling is hundreds
of bytes as well as to the incurred latency. Therefore, uplink small
data transmission (SDT) for Inactive mode UEs has been introduced
{[}5{]}. Therefore, Inactive UEs may multiplex the small uplink payload
as part of the RACH procedure, i.e., 2-step and 4-step RACH procedures.
That is, as part of the RRC configuration signaling, e.g., RRC resume
setup message. In another option, configured grant can be utilized,
where pre-configured uplink resources and transmission configuration
are adopted for quick uplink SDT. 

\subsection{Control Channel Skipping and Search Space Set Switching}

It has been understood, during release-16 evaluations, that the majority
of the UE power consumption is paid for monitoring the control PDCCH
channel during IDLE/Inactive/Connected modes {[}5{]}. This is simply
because, particularly for IDLE/Inactive PDCCH, it requires a blind
decoding operation. Therefore, Search Space Set Switching (SSSS) has
been introduced. With SSSS, several search spaces are defined with
various periodicity, monitoring duration, certain DCI formats to monitor
for (to reduce number of blind decoding's), and number of symbols.
For latency critical QoS, an SSSS with a shorter periodicity is activated;
although, for power-limited UEs, an SSSS with a larger periodicity
and shorter duration is signaled. 

SSSS procedure may result in overlapping PDCCH search spaces with
different configurations and an increased signaling overhead size.
Accordingly, it leads to a complicated network implementation. Therefore,
PDCCH control channel skipping is proposed as an alternative {[}5{]}.
With PDCCH skipping, as depicted by Fig. 6, the network dynamically
signals the power-limited UEs with an indication to safely skip monitoring
the configured PDCCH control search space for either a certain duration
or until they receive further indication to activate back PDCCH monitoring.
This way, different UEs may be configured with the same PDCCH search
space but with various monitoring patterns. Therefore, UEs, which
have been configured of PDCCH skipping, assume that no traffic will
be transmitted during the skipping duration. 

\begin{figure}
\begin{centering}
\includegraphics[scale=0.47]{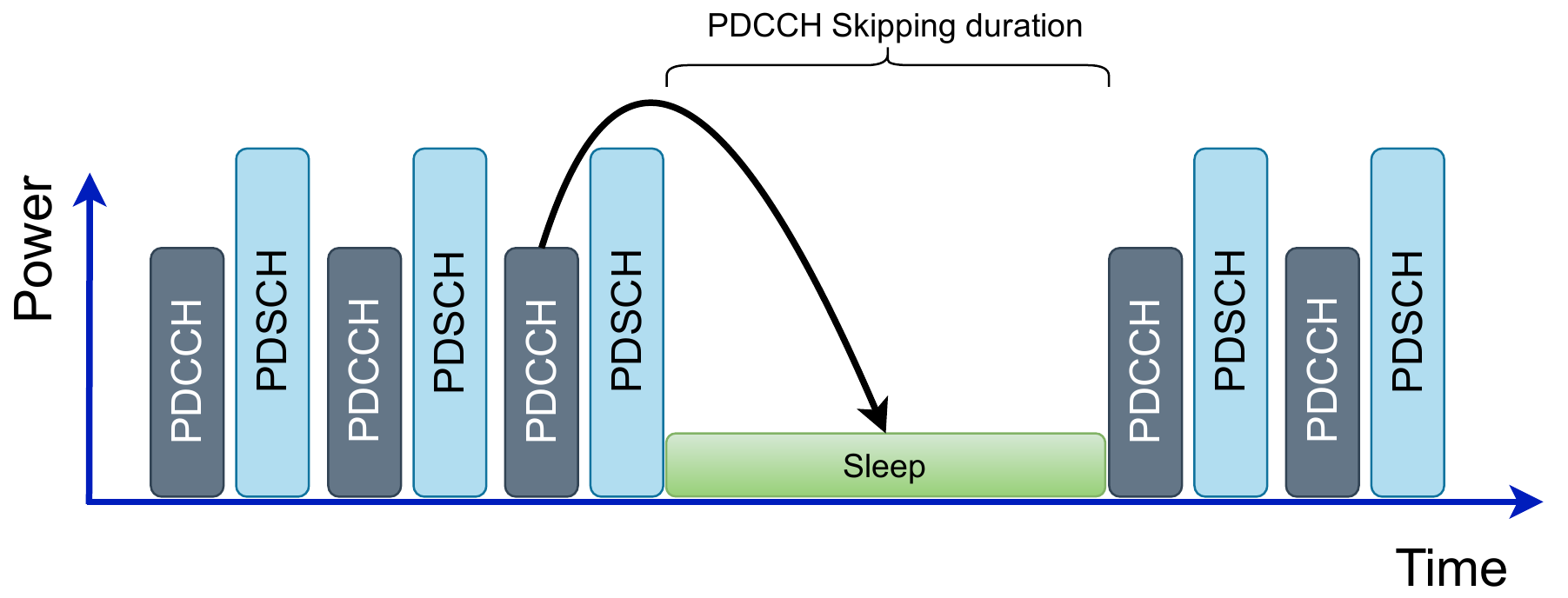}
\par\end{centering}
\centering{}Fig. 6. PDCCH skipping of 5G new radio.
\end{figure}

\section{Performance Evaluation}

Extensive system level simulations, using MATLAB, are performed to
comprehensively evaluate the achievable latency-reliability performance
for each of the considered device power saving feature. The main simulation
parameters are listed in Table I, in line with {[}10{]}. The industrial
factory deployments is considered with the 3GPP-compliant industrial
channel mode. A 4 x 4 antenna setup is adopted for the gNB and UE.
The simulation methodology, followed in this work, is inline with
{[}11{]}, where the majority of the 5G NR functionalities of Layer
1 and 2 are explicitly considered.

\begin{table}
\caption{{\small{}Simulation parameters.}}

\centering{}%
\begin{tabular}{c|c}
\hline 
Parameter & Value\tabularnewline
\hline 
Environment & 3GPP industrial factory, 18 cells\tabularnewline
\hline 
UL/DL channel bandwidth & 20 MHz, SCS = 30 KHz, FDD\tabularnewline
\hline 
Channel model & InF-DH (dense clutter and high BS)\tabularnewline
\hline 
BS and UE transmit power & BS: 30 dBm, UE: 23 dBm\tabularnewline
\hline 
Carrier frequency & 4 GHz\tabularnewline
\hline 
BS and UE heights & BS: 10 m, UE: 1.5 m\tabularnewline
\hline 
Antenna setup & 4 $\times$ 4 \tabularnewline
\hline 
Average UEs per cell & 5, 10\tabularnewline
\hline 
URLLC Traffic model & FTP3, packet size = 50 Bytes\tabularnewline
\hline 
DL/UL receiver & 	L-MMSE-IRC\tabularnewline
\hline 
\end{tabular}
\end{table}

The impact of the scheduling delays on the latency-reliability performance
is first investigated, for connected-mode UEs. Fig. 7 depicts the
complementary cumulative distribution function (CCDF) of the achievable
downlink (DL) radio latency, in a DL-only deployment. Four various
scenarios of the scheduling configurations are considered. The instant
scheduling denotes that the control PDCCH and data PDSCH channels
are assumed always available for the UEs to receive their DL grants.
The fixed offset scheduling implies configuring a minimum fixed cross-slot
scheduling among a control channel opportunity and the actual PDSCH
resources, while the dynamic offset scheduling scheme means a varying
cross-scheduling offset, different for various UEs, and all possible
offsets are larger than the configured minimum cross-slot scheduling
offset. Finally, PDCCH skipping scheme is considered for a randomly
selected sub-set of active UEs, where the skipping period is 5 slots. 

As can be observed from Fig. 7, with instant scheduling, the stringent
URLLC target of 1 ms at the $10^{-5}$ outage probability is fulfilled.
However, it comes at the expanse of the UE receivers assumed to be
always active in order to monitor the always-ON PDCCH channel, while
buffering the concurrent PDSCH resources, leading to a significant
battery power consumption. The fixed offset scheduling is implementation
and power friendly due to avoiding the buffering the PDSCH payload
which is concurrent with configured PDCCH search space(s). Although,
it obviously introduces an additional queuing delay compared to instant
scheduling. The dynamic offset scheduling exhibits \textasciitilde 180\%
increase of the outage latency, compared to the instant scheduling.
This is attributed to the additional cross-slot scheduling delay,
where the URLLC UEs on the cell-edge suffer the most, i.e., represented
by the longer tail distribution, because of the additional delay needed
for the likely multiple HARQ re-transmissions. 

The PDCCH skipping clearly degrades the outage latency performance
since UEs, during the skipping duration, are considered completely
unreachable. Therefore, with sporadic fast packet arrivals, which
arrive during the skipping period, the gNB buffers the payload, triggers
a DCI grant over the configured PDCCH search space after the skipping
period is expired, and finally, the payload is transmitted on the
allocated data resources, leading to a significant buffering delay.

\begin{figure}
\begin{centering}
\includegraphics[scale=0.55]{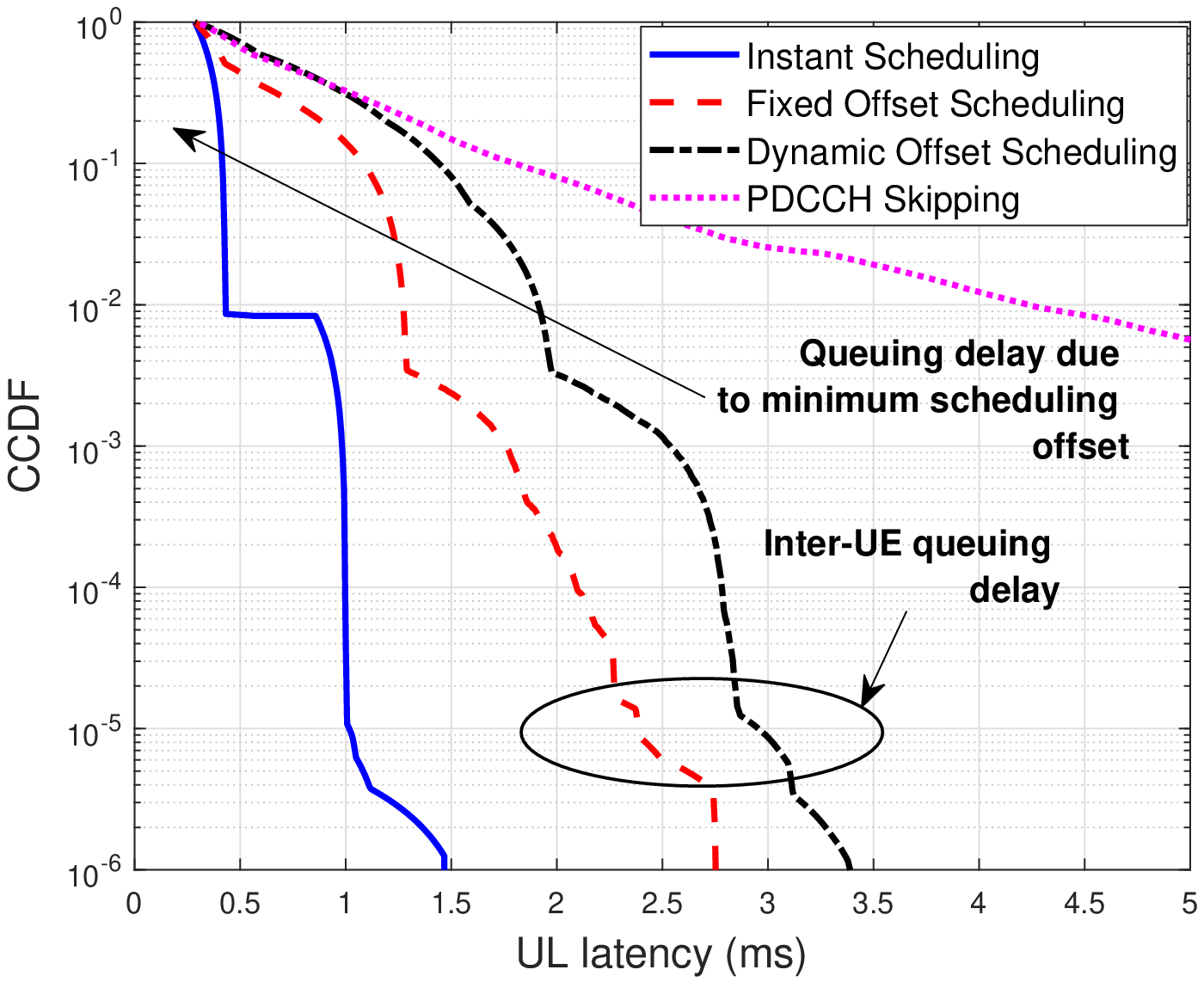}
\par\end{centering}
\centering{}Fig. 7. Downlink outage latency: impact of scheduling
delay.
\end{figure}

Similarly, the uplink (UL) outage latency performance is analyzed,
as shown by Fig. 8. Several UL transmission schemes are considered
as follows. RRC based UL transmission is adopted, where UEs, of a
newly arriving UL payload, need to establish an RRC connection prior
to the payload transmission. Furthermore, UL small data transmission
without RRC state transition is considered with configured grant (CG)
and RACH schemes, respectively. The former implies a CG is used, where
UEs transmit their new UL payload immediately on a pre-configured
set of resources, alongside with a configured preamble for the gNB
to differentiate concurrent transmissions from various UEs. The latter
option, i.e., the RACH-based SDT, implies that UEs, when a new UL
payload is available, triggers a temporary RACH procedure and multiplex
the UL payload with the RRC signaling. Accordingly, upon successful
UL payload reception at the gNB, the UE shall not transition to RRC
connected state. 

As can be seen, the RRC-based UL inflicts the worst outage latency
performance due to the requirement of the connected establishment
before payload transmission. Two scenarios of the CG based SDT are
presented. An error free CG is adopted, where the number of the CG
preambles is made sufficient enough (i.e., larger than the number
of active UL UEs), accordingly, it is enforced that various CG UEs
to select dedicated preambles, i.e., UE-specific preamble, and hence,
avoiding CG collisions. An error non-free CG implies a probability
that at least two UEs select the same preamble, hence, the CG transmissions
are collided and will be repeated over the next CG opportunity. As
depicted by Fig. 8, both CG and RACH based UL offer a similar outage
latency performance, where the latter introduces a slight increase
of the outage latency due to the processing delay of receiving and
transmitting RACH preambles before the UL payload is received as part
of the RRC resume message. It is also clear that the RACH and CG collisions
lead to a significant degradation of the outage latency, e.g., by
employing CG UL with an increasing number of active UL UEs (5 to 10
UEs), than the available CG preambles. Thus, the error non-free RACH
and CG based UL approach the outage latency performance of the conventional
RRC-based UL. 

\begin{figure}
\begin{centering}
\includegraphics[scale=0.55]{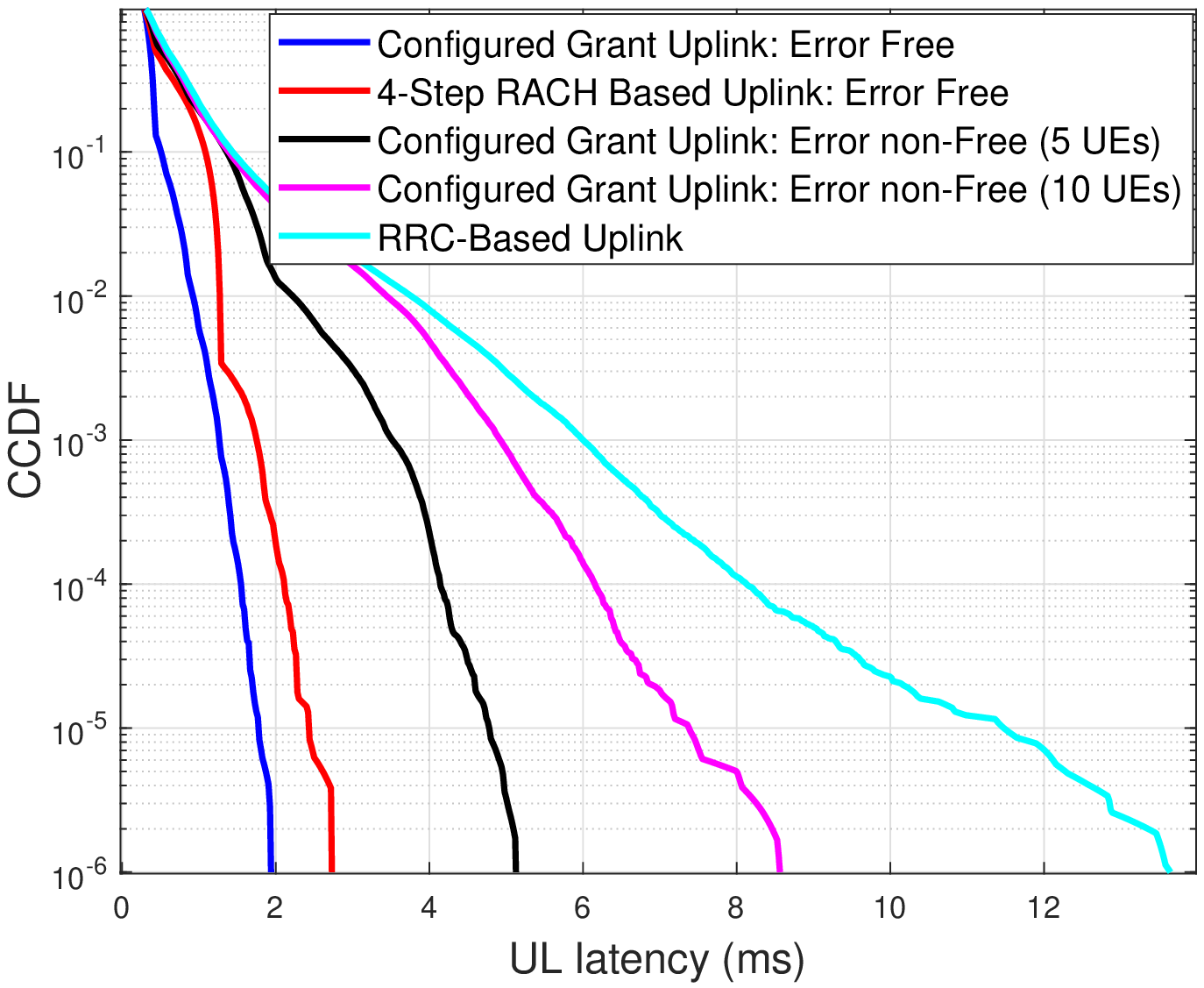}
\par\end{centering}
\centering{}Fig. 8. Uplink outage latency: impact of RRC, CG and RACH
based UL SDT.
\end{figure}

Finally, the CCDF of the wake up time for RRC IDLE/Inactive mode UEs
is depicted by Fig. 9, for various IDLE/Inactive mode power saving
features. The wake up time is defined as the total time the UE transceiver
stays active from the moment it synchronizes with the RAN interface,
detects a true paging, and successfully receive the corresponding
payload. As shown by Fig. 9, the DRX with SSB assistance for synchronization
exhibits the longest outage wake up period, i.e., up to 23 ms. This
is because UEs, and particularly cell-edge UEs of weak SINRs, need
to detect several SSBs for synchronization before the configured POs.
If UEs are not actually paged, such longer wake up period is unnecessary.
The EPI configurations are considered with three options. Common EPI
denotes an EPI transmission that is meant for all active UEs, i.e.,
a single paging UE group. Group-based EPI implies the transmission
of an EPI DCI/sequence that indicates only a single group of UEs to
be paged, i.e., multiple paging groups. IDLE RS assisted EPI means
that the transmitted EPI is prefixed by one ore more paging-specific
RSs, hence, IDLE/Inactive UEs do not need to wake up and detect the
former SSBs of a longer cycle before the POs. 

As can be clearly seen, the group-based EPI with assisting IDLE RSs
offers the best shortest wake up time needed for IDLE/Inactive UEs
to receive the respective payload. This is attributed for multiple
reasons as: (1) UEs only wake up shortly before the EPI occasion,
and synchronize with the RAN interface using the IDLE RSs, and (2)
UEs which belong to a paging group rather than the detected EPI group
shall deep sleep and skip decoding the actual PO. The IDLE DRX with
SSB assistance obviously mandate IDLE UEs to wake up for extended
periods of time for PO pre-synchronization and decoding. Moreover,
the group-based EPI decreases the UE wake up duration by \textasciitilde 50\%
compared to the common EPI transmission, i.e., a single paging group,
due to the reduced paging false alarms. 

\begin{figure}
\begin{centering}
\includegraphics[scale=0.55]{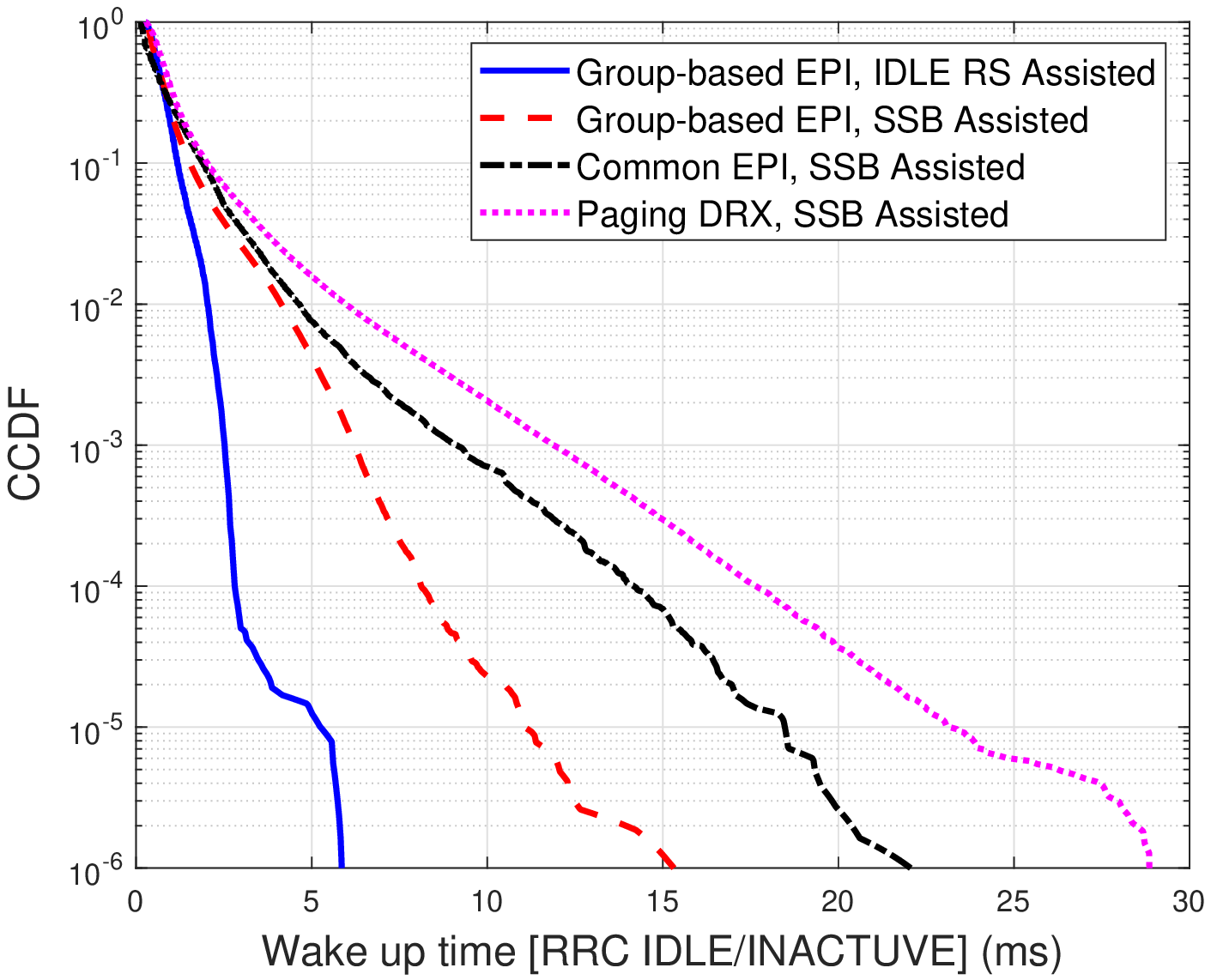}
\par\end{centering}
\centering{}Fig. 9. Wake up time: impact of 5G NR paging procedures.
\end{figure}

\section{Concluding Remarks }

In this work, a comprehensive review of the state-of-the-art 3GPP
device power saving features is presented, including paging DRX, paging
early indication, IDLE assistance reference signals, control channel
skipping, small data transmission without RRC state transition, and
cross-slot resource scheduling, respectively. Extensive system level
simulations have been performed to evaluate the impact of the various
power saving features on the achievable outage latency performance,
for latency-critical URLLC services. The major recommendations of
this work are as follows: (1) control PDCCH channel skipping is vital
for a power efficient UE operation; however, it is not proper for
latency-critical URLLC services of frequent and sporadic packet arrivals,
due to the additional buffering delays, (2) configured grant and RACH
based uplink transmissions are vital for reducing the end-to-end delay
of uplink packet transmissions. Although, preamble design and collision
avoidance measures must be guaranteed to achieve the promised latency
gain, and (3) For IDLE/Inactive mode UEs, the early paging indication,
which is scrambled with a paging-group-specific code, alongside IDLE
assisting reference signals are of a significant importance to drastically
minimize the corresponding wake up time, and the paging false alarms,
respectively.


\begin{thebibliography}{10}
\bibitem[1]{key-1} \textit{Service requirements for 5G System}, TS
22.261, V18.3.0, June 2021.

\bibitem[2]{key-2} A. A. Esswie and K. I. Pedersen, \textquotedbl Analysis
of outage latency and throughput performance in industrial factory
5G TDD deployments,\textquotedbl{} \textit{in Proc. IEEE VTC}, 2021,
pp. 1-6.

\bibitem[3]{key-3} \textit{Framework and overall objectives of the
future development of IMT for 2020 and beyond}, Recommendation ITU-R,
M.2083-0, IMT Vision.

\bibitem[4]{key-4} \textit{Study on ultra-low power wake up signal
in Rel-18}, RWS-210169, 3GPP, release-18 workshop, VIVO.

\bibitem[5]{key-5} \textit{Study on User Equipment (UE) power saving
in NR}, 3GPP, TR 38.840 (release-16), V16.0.0, June 2019.

\bibitem[6]{key-6} A. A. Esswie and K. I. Pedersen, \textquotedbl Opportunistic
spatial preemptive scheduling for URLLC and eMBB coexistence in multi-user
5G networks,\textquotedbl{} \textit{in IEEE Access}, vol. 6, pp. 38451-38463,
2018. 

\bibitem[7]{key-7} A. A. Esswie and K. I. Pedersen, \textquotedbl On
the ultra-reliable and low-latency communications in flexible TDD/FDD
5G networks,\textquotedbl{} \textit{in Proc. IEEE CCNC}, 2020, pp.
1-6.

\bibitem[8]{key-8} E. de Olivindo Cavalcante, G. Fodor, Y. C. B.
Silva, and W. C. Freitas, \textquoteleft \textquoteleft Distributed
beamforming in dynamic TDD MIMO networks with BS to BS interference
constraints,\textquoteright \textquoteright{} \textit{IEEE Wireless
Commun. Lett.}, vol. 7, no. 5, pp. 788--791, Oct. 2018.

\bibitem[9]{key-9} A. A. Esswie, K. I. Pedersen and P. E. Mogensen,
\textquotedbl Online radio pattern optimization based on dual reinforcement-learning
approach for 5G URLLC networks,\textquotedbl{} \textit{in IEEE Access},
vol. 8, pp. 132922-132936, 2020.

\bibitem[10]{key-10} K. Pedersen et al., \textquotedbl Advancements
in 5G new radio TDD cross link interference mitigation,\textquotedbl{}
\textit{in IEEE Wireless Communications}, June 2021.

\bibitem[11]{key-11} G. Pocovi, K. I. Pedersen and P. Mogensen, \textquotedbl Joint
link adaptation and scheduling for 5G ultra-reliable low-latency communications,\textquotedbl{}
\textit{in IEEE Access}, vol. 6, pp. 28912-28922, 2018.
\end{thebibliography}
\end{document}